\documentclass[11pt,a4paper,english,amssymb,nofootinbib,superscriptaddress]{revtex4}
\usepackage{lmodern}

\usepackage[T1]{fontenc}
\usepackage[latin9]{inputenc}
\usepackage{babel}
\usepackage{amsmath}
\usepackage{graphicx}
\usepackage{amssymb}
\usepackage{esint}

\usepackage{color}
\usepackage{soul}

\usepackage[unicode=true, pdfusetitle,
bookmarks=true,bookmarksnumbered=false,bookmarksopen=false,
breaklinks=false,pdfborder={0 0 1},backref=false,colorlinks=false]
{hyperref}
\def\b{\begin{equation}}
\def\e{\end{equation}}

\makeatletter
\@ifundefined{textcolor}{}
{%
 \definecolor{BLACK}{gray}{0}
 \definecolor{WHITE}{gray}{1}
 \definecolor{RED}{rgb}{1,0,0}
 \definecolor{GREEN}{rgb}{0,1,0}
 \definecolor{BLUE}{rgb}{0,0,1}
 \definecolor{CYAN}{cmyk}{1,0,0,0}
 \definecolor{MAGENTA}{cmyk}{0,1,0,0}
 \definecolor{YELLOW}{cmyk}{0,0,1,0}
 }

\usepackage{latexsym}\usepackage{bm}

\makeatother

\makeatother

\begin{document}
\title{Unitarity Problems in 3$D$ Gravity Theories}

\author{Gokhan Alkac}

\affiliation{Van Swinderen Institute for Particle Physics and Gravity,\\ University of Groningen, Nijenborgh 4, 9747 AG Groningen, The Netherlands,}

\author{Luca Basanisi}

\affiliation{Van Swinderen Institute for Particle Physics and Gravity,\\ University of Groningen, Nijenborgh 4, 9747 AG Groningen, The Netherlands,}

\author{Ercan Kilicarslan}

\affiliation{Department of Physics,\\
 Middle East Technical University, 06800, Ankara, Turkey}

\author{Bayram Tekin}

\affiliation{Department of Physics,\\
 Middle East Technical University, 06800, Ankara, Turkey}
 
\date{\today}

\begin{abstract}
We revisit the problem of the bulk-boundary unitarity clash in 2 + 1 dimensional gravity theories, which has been an obstacle in providing a viable dual two-dimensional conformal field theory for bulk gravity in anti-de Sitter ($AdS$) spacetime. Chiral gravity, which is a particular limit of cosmological topologically massive gravity (TMG), suffers from perturbative log-modes with negative energies inducing a non-unitary logarithmic boundary field theory. We show here that any $f(R)$ extension of TMG does not improve the situation. We also study the perturbative modes in  the metric formulation of minimal massive gravity-originally constructed in a first-order formulation-and find that the massive mode has again negative energy except in  the chiral limit. We comment on this issue and also discuss a possible solution to the problem of negative energy modes. In any of these theories, the infinitesimal dangerous deformations might not be integrable to full solutions; this suggests a linearization instability of AdS spacetime in the direction of the perturbative log-modes. 
\end{abstract}

\maketitle

\section{Introduction}

Three dimensional gravity is a useful theoretical laboratory to test some ideas about quantum gravity. One of the most promising ideas is the AdS/CFT duality (or more broadly holography) which, in this context, would amount to finding a gravity theory in the three dimensional bulk spacetime with a unitary, non-trivial boundary conformal field theory. But even in this simpler setting of ``flatland'', there are certain obstacles in realizing a toy model of quantum gravity. For example, cosmological Einstein's theory in $3D$ suffers from local triviality (that is, it has no propagating degrees of freedom) even though it has a healthy boundary CFT with two copies of the Virasoro algebra that have positive central charges. Since the ultimate purpose of studying these theories is to possibly learn something about the four dimensional quantum gravity, which (at least in the low energy limit) have local degrees of freedom, the local triviality of the theory makes it rather irrelevant for this purpose.

On the other hand, $3D$ cosmological Einstein's gravity has a globally non-trivial structure such as the existence of the Banados, Teitelboim, Zanelli (BTZ) black hole \cite{BTZ}. Therefore, its quantum version might still teach us a lot about the global aspects of quantum gravity. It is highly interesting that both the existence of a $3D$ black hole and the existence of a boundary CFT require a negative cosmological constant (that is the $AdS$ space). For these reasons, a lot of work has been devoted to extensions of cosmological Einstein's theory in the bulk in such a way that the extended theory has local degrees of freedom while it has the same boundary structure as Einstein's gravity with $\Lambda<0$. However, this has not been an easy task. Let us summarize what is known along these lines as there has been some recent exciting developments often followed by the realization of a potential deficiency in the proposed extensions.  
 
Among the extensions of the $3D$ Einstein's theory, Topologically Massive Gravity \cite{DJT} seems to be the most natural one with a dynamical graviton. As this theory will play a major role in the current work, let us summarize its established properties. Assuming the $3D$ Newton's constant to be positive, $G>0$, and adopting the mostly positive signature for the metric, the theory has a third derivative, parity non-invariant action given as
\begin{equation}
I=\frac{1}{16\pi G}\int d^3x \sqrt{-g}\bigg[ \sigma\bigg( R-2\Lambda\bigg)+\frac{1}{2\mu} \, \eta^{\mu \nu \alpha} \Gamma^\beta{_{\mu \sigma}} \Big (\partial_\nu \Gamma^\sigma{_{\alpha \beta}}+\frac{2}{3} \Gamma^\sigma{_{\nu \lambda}}  \Gamma^\lambda{_{\alpha \beta}}\bigg) \bigg],
\end{equation}
where $\sigma$ is dimensionless and $\mu$ has the dimension of mass, and the following properties hold:
\begin{enumerate}
\item TMG has a single massive spin-2 excitation (with $+2$ helicity for $\mu>0$) in the bulk with a mass-squared given as
\begin{equation}
m_g^2=\mu^2\sigma^2+\Lambda.
\label{gravitonmass1}
\end{equation}
In the $\Lambda \rightarrow 0$ limit, the single massive degree of freedom (DOF) remains intact, with a mass $m_g=\lvert \mu\sigma\rvert$ and a positive kinetic energy as long as $\sigma<0$, which is opposite to the Einstein's theory. Namely, if one introduces a $T_{\mu\nu}$ to the right-hand side of the field equations, the coupling between gravity and matter fields would be opposite to the four-dimensional case. This is all well-known since the original work \cite{DJT}. We should also note that in the $\sigma\rightarrow 0$ limit, there is no propagating mode: so the pure Chern-Simons theory and pure gravity do not have local DOF when considered on their own. Moreover, the pure Chern-Simons theory cannot accommodate a cosmological constant $\Lambda$ as the field equation would be inconsistent since the Cotton tensor vanishes for $AdS_3$ or any other conformally flat metric.
\item Brown-Henneaux boundary conditions for the asymptotically $AdS_3$ spacetime lead to two copies of Virasoro algebra with the left and right central charges given as
\begin{equation}
c_{L,R}=\frac{3\ell}{2G}\bigg(\sigma\mp\frac{1}{\mu\ell}\bigg),\hskip .5cm \Lambda\equiv -\frac{1}{\ell^2},
\end{equation}
where $L,R$ refer to left and right. In the $\mu\rightarrow \infty$ limit, these reduce to the ones given in \cite{Brown-Henneaux} for pure cosmological Einstein's theory with the  choice $\sigma=1$.
It was shown in \cite{chiral} that the positivity of the energy of bulk excitations with Brown-Henneaux boundary conditions requires the theory to be in the so-called chiral limit, where $\sigma^2 \mu^2 = \frac{1}{\ell^2}$. However, a closer investigation performed in \cite{Grumiller} showed that the theory in the chiral limit allows the so called log-modes as solutions which have finite but negative energy, albeit with weaker boundary conditions.
\end{enumerate}

We will now elaborate on this issue since it plays a central role in this work. The linearized fluctuations $h_{\mu\nu}\equiv g_{\mu\nu}-\bar{g}_{\mu\nu}$  around the $AdS$ background metric ($\bar{R}_{\mu\nu}=2\Lambda \bar{g}_{\mu\nu}$)  satisfy the following linearized field equations
\begin{equation}
\sigma\, {\cal{G}}^L_{\mu\nu}+\frac{1}{ \mu}C^L_{\mu\nu}=0,
\label{lineartmg}
\end{equation}
where the linearized cosmological Einstein tensor, Cotton tensor and  the scalar curvature are
\begin{equation}
\begin{aligned}
 {\cal G}^L_{\mu \nu}&=R^{L}_{\mu\nu}-\frac{1}{2} \bar{g}_{\mu\nu} R^{L}-2 \Lambda h_{\mu\nu}, \hskip .5 cm
 C^L_{\mu \nu}&=\frac{1}{2} \eta_\mu\,^{ \alpha \beta} \bar{\nabla}_\alpha {\cal{G}}^L_{\nu\beta}+\left(\mu\leftrightarrow \nu\right),\\
 R^{L}&=-\bar{\square} h+\bar{\nabla}^\mu \bar{\nabla}^\nu h_{\mu\nu}-2 \Lambda h. 
 \label{curvaturet}
 \end{aligned}
 \end{equation}
The trace of the linearized field equations (\ref{lineartmg}) demand that $R_L =0$ and one can choose the transverse-traceless gauge ($\bar{\nabla}^\mu h_{\mu\nu}=0 $ and $h=0$) which is compatible with the trace equation. This choice leads to a rather remarkable simplification of the field equation: it splits into three factors as \cite{chiral} 
\begin{equation}
\Big ( \mathcal{D}^L  \mathcal{D}^R   \mathcal{D}^{m_g}    h \Big )_{\mu \nu}=0,
\label{product}	
\end{equation}
with the three first order operators defined as
\begin{eqnarray}
 ( \mathcal{D}^{L/R})_\mu\,^\nu \equiv \delta_\mu^\nu \pm \ell  \,\eta_{\mu}\,^{\alpha \nu} \bar\nabla_\alpha, \nonumber 
\hskip .5 cm ( \mathcal{D}^{m_g})_\mu\,^\nu \equiv \delta_\mu^\nu + \frac{1}{\mu\sigma} \eta_{\mu}\,^{\alpha \nu} \bar\nabla_\alpha.
\end{eqnarray}
These operators are mutually commuting save at two degeneration points $\sigma^2\mu^2=\frac{1}{\ell^2}$ (which we shall discuss separately below, as they turn out to be the only relevant points in the theory). Away from the degeneration points, the cubic derivative field equations split into three individual first derivative equations
\begin{equation}
 ( \mathcal{D}^L h^L )_{\mu \nu}=0,\hskip .5 cm  ( \mathcal{D}^R h^R  )_{\mu \nu}=0,  \hskip .5 cm ( \mathcal{D}^{m_g} h^{m_g} )_{\mu \nu}=0,
\end{equation} 
with the most general solution to (\ref{lineartmg}) given as a sum of all possible solutions to these equations $h_{\mu\nu}=h^L_{\mu\nu}+h^R_{\mu\nu}+h^{m_g}_{\mu\nu}$ consistent with the boundary conditions, which arise from physical requirements such as the finiteness of the energy and the well-posedness of the variational problem. Hence, one must compute the energies of the excitations and, for this purpose, one needs the ${\cal{O}}(h^2)$ action yielding (\ref{product}) upon variation. This can be easily found to be 
\begin{equation}
I=\frac{\sigma}{64\pi G\Lambda}\int d^3x \sqrt{-\bar{g}} h^{\mu\nu}\Big ( \mathcal{D}^L  \mathcal{D}^R   \mathcal{D}^{m_g}    h \Big )_{\mu \nu}.
\end{equation}
Using the Ostragradsky method, one finds the energies of these excitations for the left-right modes as
\begin{equation}
E_{L/R} = - \frac{c_{L/R}}{48\pi\ell}  \int d^2 x \, \sqrt{-\bar{g}} \,\bar{\nabla}^0  h_{L/R}^{\mu \nu} \, \partial_t h^{L/R}_{\mu \nu}, 
\label{energy1}
\end{equation}
while for the massive mode, the energy reads
\begin{equation}
E_{m_g} = \frac{m_g^2}{64\pi\mu G}   \int d^2 x \, \sqrt{-\bar{g}} \, \eta _\alpha\,^{0\mu} h_{m_g}^{\alpha \nu}\, \partial_t h^{m_g}_{\mu \nu}.\label{energy2}
\end{equation}
One must compute these integrals for all the ${\it normalizable}$ and viable (in the sense discussed above) solutions of the theory. So this is really a non-trivial task but it was elegantly done in \cite{chiral} where the solutions were obtained from the fact that they furnish a representation of the $AdS_3$ algebra that is $SL(2,R) \times SL(2,R)$.  Primary states are found as usual and their descendants are computed. 

To summarize, we know all the solutions of the theory and, for these solutions, the integrands above are ${\it negative}$. Therefore, to make the energies positive one must have positive or vanishing central charges and set the mass of the bulk excitation to zero, {\it{i.e}} $m_g = 0$.  The  latter gives  the constraint $\sigma^2\mu^2=\frac{1}{\ell^2}$. Choosing $\mu>0$ without loss of generality, this leads to $c_L=0$ and $c_R=\frac{3}{\mu G}$. For the massive mode to be non-tachyonic, it has to satisfy the Breitenlohner-Frieedman (BF) bound \cite{BF} $m_g^2\geq-\frac{1}{\ell^2}$, which it does. 

An additional constraint arises from the energy of the BTZ black hole. Being locally an Einstein manifold, the BTZ solution of any kind (rotating or non-rotating) solves the TMG equation without a change in the metric. Considering the non-rotating case for which the Chern-Simons term does not contribute to the energy \cite{Deser-Tekin}, one obtains $E_{BTZ}=M\sigma$ up to a positive multiplicative constant which we take it to be one.
Therefore, it seems that choosing $\sigma>0$ ostensibly makes the theory bulk and boundary unitary with only a right moving CFT with one copy of Virasoro algebra on the boundary. This is the ${\it chiral}$ gravity proposed in \cite{chiral}. 

However, to satisfy all the bulk and boundary unitarity constraints together, we have to consider exactly the, up to now deterred, {\it{degenerate limit}} for which the massive-mode operator $\mathcal{D}^{m_g}$ becomes the same as the left-moving mode operator $\mathcal{D}^{L}$ or the right-moving mode operator $\mathcal{D}^{R}$, rendering the theory irreducibly higher derivative. This means that we have not done what we promised to do: take all the viable solutions of the linearized theory. Namely, now we have additional solutions  which are not covered by our earlier ``generic'' solution. Considering $\mu\sigma=\frac{1}{\ell}$, these solutions do not satisfy $(\mathcal{D}^{L}h)_{\mu\nu}=0$, but satisfy the quadratic equation $(\mathcal{D}^{L}\mathcal{D}^{L}h)_{\mu\nu}=0$. When these solutions are explicitly constructed, one observes that  they are of the so-called log-mode form \cite{ Grumiller} and do not obey the Brown-Henneaux boundary conditions. Notwithstanding, these solutions still satisfy the requirement that spacetime is asymptotically $AdS_3$ with a weaker decaying behavior and a linearly-growing time profile. Moreover, they have a well-posed variational formulation as shown in \cite{Grumiller} and a finite boundary stress tensor. These solutions have finite, yet negative energy and hence they cannot be eliminated without further constraints. They are the ghosts of the chiral gravity, which seem to ruin the possibility of defining a quantum version of the bulk theory. 

We now explain how the chiral theory might escape these potentially problematic perturbative log-modes. It is true that these modes survive and ruin the theory in this naive perturbative setting, but one has to be careful when applying perturbative techniques in a non-linear theory such as the one studied here. For the perturbation theory to make sense in such a set-up, the perturbative solution should be {\it {integrable}} to a full solution.  Namely,  in the theory, there must exist a solution metric $g_{\mu\nu}(s)$, with $s$  being a deformation parameter, for which we have $g_{\mu\nu}(0) = \bar g_{\mu\nu}$ where the latter is the $AdS$ metric. An infinitesimal deformation of this background is given as
	\begin{equation}
	h_{\mu\nu}\equiv\frac{d}{ds}g_{\mu\nu}( s)\biggr\rvert_{s=0}.
	\end{equation} 
If the infinitesimal deformation $h_{\mu \nu}$ (obtained as a solution to the linearized field equations)  does not satisfy this property, even if it solves the linearized equation, it cannot be integrable to a full solution, i.e. there does not exist an exact solution $g_{\mu\nu}( s)$ whose linearized version yields $h_{\mu\nu}$ about the point $\bar{g}_{\mu\nu}$. This implies that the background metric has a linearization instability \cite{Fischer1980} and hence perturbation theory about that background is inconsistent at this order. It is known that in such a case higher order terms in the perturbative expansion bring in constraints on the linear order solution. More properly, at the second order the "Taub charges" (an integral whose integrand is quadratic in the first order solution) must vanish.  The linearization instability appears in various contexts in gravity. For example, for compact Cauchy hypersurfaces, one has linearization instability in pure General Relativity \cite{Deser1973}. As another example, the spacetime {\it effectively} has compact Cauchy hypersurfaces for theories with pure higher curvature terms which require stronger decaying conditions \cite{Strominger_positive}. As a solution to the problem of log-modes in chiral TMG, it was proposed in \cite{Maloney, Carlip} that the log-modes could also be eliminated as there is a linearization instability in chiral TMG. In other words, there are constraints on the linearized solutions coming at the second order. 

There have been other proposals for a viable theory in $3D$. New Massive Gravity (NMG) \cite{NMG} is one such example: instead of the Chern-Simons term, one augments the cosmological Einstein theory with the judiciously chosen quadratic combination $ K =R_{\mu\nu}^2-\frac{3}{8}R^2$ and the resulting theory has a massive spin-2 excitation with both of the helicities present.  However, also this theory is either bulk or boundary unitary but not both. In fact,  it was shown in \cite{Sisman1} that \textit{no extension} of Einstein's gravity in $3D$ that has the same particle content as NMG can be free of the bulk-boundary unitarity conflict. For example, cubic, quartic, and higher order extensions of NMG that are consistent with the bulk AdS/CFT constraints, such as the existence of a $c$-function, turned out to be non-unitary on the boundary \cite{Sinha,Paulos}. Infinite order extension of NMG in the form of a Born-Infeld gravity also fails to be unitary on the boundary \cite{Gullu, Gullu2}, even though it has all the nice bulk properties and the highly desired property of having a unique vacuum a property that neither MMG nor its finite order extensions have. For further work on these theories see \cite{Naseh,Ohta}.  

After all these vigorous attempts, one might wonder if a $ 3D$ gravity that is amenable to holography or AdS/CFT arguments exists . Minimal Massive Gravity (MMG) was proposed as such a theory in \cite{mmg}, where it is obtained from an action in the first-order formulation with auxiliary 1-form fields. However, it is still possible to write down its field equations with the metric being the only dynamical field, where the TMG's field equations are modified with a {\it part } of the field equations coming from the purely quadratic $K$-gravity \cite{KGravity} whose Lagrangian is noted above.  Namely, consider the quadratic curvature invariant $K$, its variation leads to the tensor $K_{\mu\nu} =  J_{\mu\nu} + H_{\mu\nu}$, with  $J_{\mu\nu}$ and $H_{\mu\nu}$ given as \cite{MMG2}
\begin{equation}
 J^{\mu\nu}\equiv \frac{1}{2}\eta^{\mu\rho\sigma}\eta^{\nu\alpha\beta}S_{\rho\alpha}S_{\sigma\beta},\hskip .5 cm
  H^{ \mu \nu }  \equiv \frac{1}{2}\eta^{\mu \alpha \beta }\nabla_\alpha C^\nu_\beta +  \frac{1}{2}\eta^{\nu \alpha \beta }\nabla_\alpha C^\mu_\beta.
  \label{jtensor}
\end{equation}
If the field equations of TMG are  extended with $J_{\mu\nu}$, the particle content remains intact. On the other hand, if the field equations of TMG are extended with $H_{\mu\nu}$, the particle spectrum doubles: the theory has a massive spin-2 excitation with two helicities that have different masses and an ostensibly improved boundary behavior \cite{MMG2}. These two latter theories do not come from covariant actions where the metric field is the only dynamical variable. The consistency of the field equations as classical equations are satisfied  with the help of on-shell Bianchi identities. Both theories were claimed to have an improved boundary behavior. However, we will investigate the metric perturbations for the case of MMG and show that one must go to the chiral limit of the theory for the positivity of the energy of the massive mode.  Unfortunately, exactly at that point, the log-modes arise as perturbative solutions, which seem to ruin the unitarity again.

 We will also show that there are other  log-modes which are {\it exact} solutions of all these theories (TMG, NMG, MMG) considered here. These log-modes appear exactly at the chiral point, but they are of the {\it{wave-type}} coming with an arbitrary function. Additionally, they are constructed in the Poincar\'e patch as opposed to the global coordinates, where the perturbative log-mode becomes manifest. Since the solutions include an arbitrary function and a modified boundary behavior, they do not pose any threat to the theory.

At this moment, it is clear that one of the potential candidates of a viable unitary theory is the chiral gravity granted that the linearization instability removes the ghost modes. Still, the question that needs to be addressed is: can one deform/extend TMG in such a way the chiral limit is avoided and the log-modes with negative energy disappear? Even if the answer is no, the linearization instability might occur in any of these theories since they differ at the non-linear level. To answer this question, we extend TMG with $f(R)$-type terms. We will see that, in these theories, one still must go to the chiral gravity limit and hence the perturbative log-modes are still dangerous. As mentioned earlier, we will also show that, contrary to earlier claims, MMG does not seem to provide any improvement as long as the energy of bulk excitations are concerned and the theory should be considered only in the chiral limit.

\section{Topologically Extended $f(R)$ Gravity}
We shall study a generic $f(R)$ extension of TMG but, as it will serve as the basis for the generic theory, first let us work with the quadratic extension defined by the action
\begin{equation}
I=\frac{1}{16\pi G}\int d^3x \sqrt{-g}\bigg[\sigma \bigg(R-2\Lambda_0\bigg)+\alpha R^2+\frac{1}{2\mu} \, \eta^{\mu \nu \alpha} \Gamma^\beta{_{\mu \sigma}} \Big (\partial_\nu \Gamma^\sigma{_{\alpha \beta}}+\frac{2}{3} \Gamma^\sigma{_{\nu \lambda}}  \Gamma^\lambda{_{\alpha \beta}}\bigg) \bigg],
\label{genaction}
\end{equation}
whose source-free field equations read
\begin{equation}
\sigma \bigg(R_{\mu\nu}-\frac{1}{2}g_{\mu\nu}R+\Lambda_0 g_{\mu\nu}\bigg)+2\alpha R\bigg(R_{\mu\nu}-\frac{1}{4}g_{\mu\nu}R\bigg)
+2\alpha\bigg(g_{\mu\nu}\square-\nabla_\mu\nabla_\nu\bigg)R+\frac{1}{\mu}C_{\mu\nu}=0.
\label{fieldeqns1}
\end{equation}
The theory has two maximally symmetric vacua with the effective cosmological constants given as
\begin{equation}
\Lambda_{\pm}=\frac{\sigma}{12 \alpha}\bigg(1\pm \sqrt{1-\frac{24\alpha\Lambda_0 }{\sigma}}\bigg),
\label{vacuum}
\end{equation}
as long as $1-\frac{24\alpha\Lambda_0 }{\sigma}\geq 0$; when the bound is saturated the two vacua merge. In the limit $\alpha \rightarrow 0$, the vacuum with $\Lambda_+$ disappears and one recovers TMG.
Linearization of the field equations about one of these vacua yields
\begin{equation}
\tilde {\sigma} \,{\cal{G}}^L_{\mu\nu}
+2\alpha \bigg(\bar{g}_{\mu\nu}\bar{\square}-\bar{\nabla}_\mu\bar{\nabla}_\nu+2\Lambda\bar{g}_{\mu\nu}\bigg)R_L+\frac{1}{\mu}C^L_{\mu\nu}=0,
\label{lineareq}
\end{equation}
where $\tilde {\sigma}=\sigma+12\Lambda\alpha$. To work out the particle content of the theory, let us take the trace of the linearized field equations which yields a massive scalar 
equation 
\begin{equation}
\bigg(\bar{\square}-m_s^2\bigg)R^L=0,\hskip 1 cm m_s^2=\frac{\tilde{\sigma}}{8\alpha}-3\Lambda.
\label{scalarmass}
\end{equation}
This mode decouples as the theory reduces to the TMG. Another point to note is that, unlike the TMG case, one {\it cannot} choose a gauge which makes the gauge-invariant dynamical quantity $R_L$ vanish. The transverse-traceless gauge that we employed above for TMG, which significantly simplified the computations, is not consistent in this theory. To decouple the massive spin-$0$ graviton and the massive spin-$2$ graviton, let us employ the methods used in \cite{Tekin:2016}. The ${\cal{O}}(h^2)$ action yielding the linearized field equations (\ref{lineareq}) is
\begin{equation}
I_{{\cal{O}}(h^2)}=-\frac{1}{32\pi G}\int d^3x \sqrt{-\bar{g}}\bigg(\tilde{\sigma}h^{\mu\nu} {\cal {G}}^L_{\mu\nu}-2\alpha R_L^2+\frac{1}{\mu}h^{\mu\nu} C^L_{\mu\nu}\bigg),
\end{equation}
which can be recast with the help of an auxiliary field $\varphi$ as
 \begin{equation}
I(h,\varphi)=-\frac{1}{32\pi G}\int d^3x \sqrt{-\bar{g}}\bigg(\tilde{\sigma}h^{\mu\nu} {\cal {G}}^L_{\mu\nu}-4\alpha\varphi R_L+2\alpha\varphi^2+\frac{1}{\mu}h^{\mu\nu} C^L_{\mu\nu}\bigg).
\end{equation}
To decouple the $\varphi$ field from the spin-$2$ field, let us make the redefinition $h_{\mu\nu}\equiv f_{\mu\nu}-\frac{4\alpha}{\tilde{\sigma}}\bar{g}_{\mu\nu}\varphi$, which leaves the Cotton part intact and reduces the action to 
\begin{equation}
I_{{\cal{O}}(h^2)}=-\frac{1}{32\pi G}\int d^3x \sqrt{-\bar{g}}\bigg(\tilde{\sigma}f^{\mu\nu} {\cal {G}}^L_{\mu\nu}(f)-\frac{16\alpha^2}{\tilde{\sigma}}\varphi\bigg(\bar{\square}+3\Lambda-\frac{\tilde{\sigma}}{8\alpha}\bigg)\varphi +\frac{1}{\mu}f^{\mu\nu} C^L_{\mu\nu}(f)\bigg),
\label{linaction1}
\end{equation}
from which the mass of the scalar mode given in (\ref{scalarmass}) follows.  It is important to note that the scalar field comes with the correct non-ghost kinetic energy for $\tilde{\sigma}>0$. 
In this case, one can bring the Lagrangian of the scalar field to the canonical form by simply rescaling $\varphi$ as $\tilde{\varphi}=\frac{4\alpha}{\sqrt{\tilde{\sigma}}}\varphi$. Once the scalar field is decoupled, the rest of the action is just the TMG action with a modified Newton's constant ($\sigma \rightarrow \tilde{\sigma}$), thus the mass of the single spin-2 mode is 
\begin{equation}
m_g^2=\mu^2(\sigma+12\Lambda\alpha)^2+\Lambda.
\label{gravitonmass}
\end{equation}
From the ${\cal{O}}(h^2)$ action (\ref{linaction1}), it is possible to deduce that the theory has a ghost in the flat space limit ($\Lambda\rightarrow 0$). In \cite{Townsend, Gullu2010}, where $h_{\mu\nu}$ is decomposed into its irreducible parts in a gauge-invariant way, it was shown that the massive spin-$2$ mode is a ghost if $\tilde{\sigma}>0$. On the other hand, if $\tilde{\sigma}<0$, the massive spin-$0$ is a ghost. Therefore,  there is no way to avoid the ghost mode in flat backgrounds. As we will show, the problem is cured in the $AdS$ backgrounds. 

Before that, we first carry out this computation in another way, which yields a good insight of how the full theory reduces to the scalar matter-coupled TMG.  For this purpose, let us consider the generic perturbation as
\begin{equation}
h_{\mu\nu}=h^{TT}_{\mu\nu}+\bar{\nabla}_{(\mu} V_{\nu)}+\bar{\nabla}_\mu\bar{\nabla}_\nu\phi+\bar{g}_{\mu\nu}\psi,
\end{equation}
with $\bar{\nabla}^\mu h^{TT}_{\mu\nu}=0$, $h^{TT}\equiv \bar{g}^{\mu\nu}h^{TT}_{\mu\nu}=0$ and $\bar{\nabla}^\mu V_\mu=0$. This decomposition leads to 
\begin{equation}
\begin{aligned}
& {\cal {G}}^L_{\mu\nu}=-\frac{1}{2}\bigg(\bar{\square} -2\Lambda\bigg)h^{TT}_{\mu\nu}+\frac{1}{2}\bigg(\bar{g}_{\mu\nu}\bar{\square} -\bar{\nabla}_\mu\bar{\nabla}_\nu+2\Lambda\bar{g}_{\mu\nu}\bigg)\psi\\
&R_L=-2\bigg(\bar{\square} +3\Lambda\bigg)\psi.
\end{aligned}
\end{equation}
At this stage, one must note that there is a constraint on $\psi$ as $\left(\bar{\square} +3\Lambda\right)\psi\neq 0$, otherwise the dynamical field is killed.
The linearized Cotton tensor is susceptible only to  the transverse-traceless part and hence $C^L_{\mu\nu}(h_{\alpha\beta})=C^L_{\mu\nu}(h^{TT}_{\alpha\beta})$. Then  (\ref{lineareq})
 becomes 
 \begin{equation}
-\frac{\tilde{\sigma}}{2}\bigg(\bar{\square} -2\Lambda\bigg)h^{TT}_{\mu\nu}-\frac{1}{2\mu}\eta_\mu\,^{\alpha\beta}\bar{\nabla}_\alpha\bigg(\bar{\square} -2\Lambda\bigg)h^{TT}_{\beta\nu}
=T^{TT}_{\mu\nu},
\label{TTeqn}
\end{equation}
where  the scalar mode appears as a source term with the traceless energy-momentum tensor 
\begin{equation}
T^{TT}_{\mu\nu}=4\alpha\bigg( \frac{\bar{g}_{\mu\nu}}{3}\bar{\square} -\bar{\nabla}_\mu\bar{\nabla}_\nu\bigg)\bigg(\bar{\square} -m_s^2\bigg)\psi.
\label{TTtensor}
\end{equation}
In (\ref{TTeqn}) the Cotton part does not look explicitly symmetric, but it is nevertheless symmetric as can be easily checked.
Since $T^{TT}_{\mu\nu}$ vanishes on-shell, the  massive spin-2 graviton equation reduces exactly to the TMG form with a modified mass given before (\ref{gravitonmass}).

We can now summarize the effects of adding the $\alpha R^2$ term to the TMG action:
the effective cosmological constant changes, a massive spin-$0$ degree of freedom is introduced, and the mass of the single massive spin-$2$ particle is modified non-trivially from that of TMG.
All of these modifications also affect the boundary theory. However, the analysis of the massive spin-2 mode reduces to the pure TMG case with the identification ($\sigma \rightarrow \tilde{\sigma}$), as it should be apparent from the ${\cal{O}}(h^2)$ action (\ref{linaction1}). With this information at hand, let us now restudy the bulk and boundary unitary.  First of all, we have the massive spin-$2$ particle which should obey the BF condition
\begin{equation}
m_g^2=\mu^2\bigg(\sigma+12\Lambda\alpha\bigg)^2+\Lambda\geq \Lambda,
\end{equation}
which is automatically satisfied. Furthermore, expressions for the excitations' energies (\ref{energy1}-\ref{energy2}) remain intact, but in the explicit solutions of the field equations, compared to the TMG case that has the $\sigma\mu$ combination, we now have the modified value $\tilde{\sigma}\mu$. Hence, the conditions on the energy of the excitations can be simply recast as
\begin{equation}
c_{L,R}=\frac{3\ell}{2G}\bigg(\sigma-\frac{12\alpha}{\ell^2}\mp\frac{1}{\mu\ell}\bigg)\geq 0, \hskip 1.5 cm \mu^2\ell^2\bigg(\sigma-\frac{12\alpha}{\ell^2}\bigg)^2\leq 1,
\end{equation}
the latter coming from the energy of the massive spin-$2$ mode. Just like in TMG, these three conditions are compatible only in the chiral gravity limit for which one has
\begin{equation}
\mu\ell\bigg(\sigma-\frac{12\alpha}{\ell^2}\bigg)=1,
\end{equation} 
which yields the vanishing of the bulk massive spin-$2$ and left-moving modes since $c_L=0$. We also have a spin-$0$ mode, which must be a non-ghost and non-tachyonic; we thus require 
\begin{equation}
m_s^2=\frac{\sigma}{8\alpha}-\frac{3\Lambda}{2}\geq 0.
\end{equation} 

Since the BTZ black hole is locally $AdS_3$,
	it also solves our theory defined by (\ref{fieldeqns1}). The positivity of its energy should also be imposed as a consistency condition
	\begin{equation}
	E_{BTZ}=M\bigg(\sigma+12\Lambda\alpha\bigg) > 0,
	\end{equation}
	where we again dropped an irrelevant (positive) multiplicative constant.

 Finally, these conditions must be compatible with the existence of an $AdS$ vacuum with the value given as (\ref{vacuum}). One can show that there are regions in the $\{\sigma,\mu,\Lambda_0,\alpha\}$ parameter space that satisfy all the  constraints. A detailed analysis of the total parameter space is not necessary for our purposes. Let us consider a specific example, choosing the $+$ branch in (\ref{vacuum}) and setting $\sigma=1$, one finds 
 \begin{equation}
 \Lambda_0=-2\mu^2, \hskip .5 cm \Lambda=-4\mu^2,\hskip .5 cm \alpha=-\frac{1}{48\mu^2},\hskip .5 cm m_s^2=0,
 \end{equation}
in addition to $m_g^2=0$, which is already the case in the chiral limit.

Just like TMG and MMG, $R^2$ extension of TMG or any other $f(R)$ extension, to be discussed below, have two different kinds of log-modes at the chiral point. The exact wave-like log-mode looks too complicated in the global $AdS_3$ coordinates, hence we will depict it  in the Poincar\'e patch in the next section. The more dangerous perturbative log-mode can be written in the global coordinates and its problematic properties were studied in \cite{Grumiller}. 

Now, consider the generic $f(R)$ gravity coupled to the Chern-Simons theory with the action
\begin{equation}
I=\frac{1}{16\pi G}\int d^3x \sqrt{-g}\bigg[f(R)+\frac{1}{2\mu} \, \eta^{\mu \nu \alpha} \Gamma^\beta{_{\mu \sigma}} \Big (\partial_\nu \Gamma^\sigma{_{\alpha \beta}}+\frac{2}{3} \Gamma^\sigma{_{\nu \lambda}}  \Gamma^\lambda{_{\alpha \beta}}\bigg)\bigg],
\end{equation}
The vacuum and the mass content of this theory can be obtained from the so-called equivalent quadratic action \cite{Gullu:2010} defined by the Lagrangian density
\begin{equation}
f_{quad-equal}=\sigma (R-2\Lambda_0)+\alpha R^2,
\end{equation}
where we did not include the Chern-Simons part, which can be added later. The parameters $\sigma$, $\Lambda_0$, and $\alpha$ can be found in terms of the function $f(R)$ and its derivatives as
\begin{equation}
\sigma=f_R(\bar{R})-6\Lambda f_{RR}(\bar{R}), \hskip 1 cm \sigma \Lambda_0=-\frac{f(\bar{R})}{2}+3\Lambda f_R(\bar{R})-9\Lambda^2  f_{RR}(\bar{R}),
\end{equation}
where $ f_{R}(\bar{R})=\frac{\partial f}{\partial R}\vert_{\bar{R}}$ {\it etc}. Finally, the vacuum equation $\sigma (\Lambda-\Lambda_0)-6\alpha \Lambda^2=0$ should also be satisfied. This analysis shows that the quadratic theory we studied above is a template for all $f(R)$ theories: as far as their particle content and vacua are considered, these theories boil down to the quadratic theory that we discussed above. So, given the explicit form of the Lagrangian, one can work out the constraints coming from the bulk and boundary unitarity.

Using the equivalence of $f(R)$ gravity and the scalar-tensor theory \cite{DeFelice}, $f(R)$ extended TMG can be mapped to a scalar-matter coupled TMG with a modified Newton's constant. The action (\ref{genaction}) can be written up to certain field redefinitions and boundary terms as
	\begin{equation}
	I=\frac{1}{16\pi G}\int d^3 x \sqrt{-g}\bigg(\beta R-\frac{1}{2}\partial_\alpha \omega \partial^\alpha \omega-V(\omega)+\frac{1}{2\mu} \, \eta^{\mu \nu \alpha} \Gamma^\beta{_{\mu \sigma}} \Big (\partial_\nu \Gamma^\sigma{_{\alpha \beta}}+\frac{2}{3} \Gamma^\sigma{_{\nu \lambda}}  \Gamma^\lambda{_{\alpha \beta}}\bigg),
	\end{equation}
	where the potential $V(\omega)$ is given by
	\begin{equation}
	V(\omega)=e^{-\frac{3\omega}{2\sqrt{\beta}}}\bigg(2\sigma\Lambda_0+\frac{1}{4\alpha}(\beta e^{\frac{\omega}{2\sqrt{\beta}}}-\sigma)^2\bigg),
	\end{equation}
	and $\beta$ is a positive constant modifying the Newton's constant. 

Having established the equivalence of our model with TMG coupled to  the scalar matter, let us briefly revisit the Birkhoff theorem for TMG, which was formulated in \cite{Aliev2,Cavaglia:1999si}. It states that the most general solution of TMG with the topology $\Sigma_2 \times S$ is static and locally Einstein trivial, {\it{i.e.}} the Cotton tensor vanishes for the solution. As discussed in \cite{Cavaglia:1999si}, the conclusion does not change for the scalar matter coupled theory since $T{^u}{_\phi}=0$ and $T{^v}{_\phi}=0$, where $u$ and $v$ are light-cone coordinates on $\Sigma_2$, and $T{^\mu}{_\nu}$ is the stress-energy tensor of the matter coupled theory. From the equivalence, the theorem is also valid for the $f(R)$-modified TMG, which this time follows from the fact that $\mathcal{A}{^u}{_\phi}=0$ and $\mathcal{A}{^v}{_\phi}=0$ where $\mathcal{A}{^\mu}{_\nu}$ is the contribution of the $f(R)$-term to field equations. Therefore, to get contributions from the Chern-Simons term, one needs to consider the case with a twist, namely, the Killing vector should not be hypersurface orthogonal {\cite{Aliev2}}. 

\section{Minimal Massive Gravity at the Chiral limit}

In \cite{Tekincharge, Alishahiha}, the chiral limit of MMG was studied but it was not realized that the theory is well-defined only in the chiral limit once all the unitarity constraints are taken into account. In this section, we will briefly discuss the analysis for MMG in the metric-formulation and prove this result.

The source-free field equations of MMG are given as \cite{mmg}
\begin{equation}
\sigma G_{\mu\nu} +\Lambda_0\,  g_{\mu\nu} + \frac{1}{\mu} C_{\mu\nu} + \frac{\gamma}{\mu^2} J_{\mu\nu} = 0\, , 
\label{mmg_denk}
\end{equation}
where the $J_{\mu\nu}$ tensor given in (\ref{jtensor}) explicitly reads  
\begin{equation}
J_{\mu \nu} = G_\mu^\rho G_{\rho \nu} - \frac{1}{2} g_{\mu \nu} G_{\rho \sigma} G^{\rho \sigma} + \frac{1}{4}G_{\mu \nu} R + \frac{1}{16} g_{\mu \nu} R^2, \label{J-ten}
\end{equation}
which is not covariant  divergence-free but the field equations are nevertheless consistent on-shell.

We now take a different approach than \cite{mmg} and study the properties of the theory starting from its metric field equations (\ref{mmg_denk}). Linearizing the field equations around one of its two vacua yields
\begin{equation}
	\bigg(\sigma -  \frac{\gamma \Lambda }{2 \mu^2}\bigg)\,{\cal G}_{\mu\nu}^L + \frac{1}{\mu} C^L_{\mu \nu}
	= 0,
	\label{lin_ark}
	\end{equation}
which is nothing but the linearized TMG with a modified Newton's constant. Hence, just like TMG, the theory has a simple massive bulk parity-non invariant spin-2 graviton with a mass-squared 
\begin{equation}
m_{\mbox{g}}^2=   \mu^2 \left (\sigma -  \frac{\gamma \Lambda }{2 \mu^2} \right)^2 + \Lambda.
\end{equation}
For the computation of the central charges, we use the method of \cite{Liu:2009pha}, where the central charges are obtained as  conserved charges corresponding to the asymptotic Killing vectors where the linearized equations and the assumption of Brown-Henneaux boundary conditions are enough for the computation. It suggests that the  the left and right central charges of the boundary CFT are
\begin{equation}
c_L = \frac{ 3 \ell}{2 G_3} \left ( \sigma + \frac{ \gamma}{2 \mu^2 \ell^2 } - \frac{1}{\mu \ell} \right ),  \hskip 1 cm 
c_R = \frac{ 3 \ell}{2 G_3} \left ( \sigma + \frac{ \gamma}{2 \mu^2 \ell^2 } + \frac{1}{\mu \ell} \right ). \label{central}
\end{equation}

The above discussion shows that the ensuing discussion will not be different from that of TMG, but for the sake of completeness, let us give a little more detail on the energies of the excitations following \cite{chiral}.  For this purpose one needs to find the ${\cal{O}}(h^2)$ action yielding the linearized field equations (\ref{mmg_denk}), which up to a boundary term reads as 
\begin{equation}
\begin{aligned}
I&=\frac{1}{64\pi G}\int d^3x
 \sqrt{-\bar{g}}\bigg(-(\sigma +  \frac{\gamma  }{2 \mu^2\ell^2})\bar{\nabla}^\rho h^{\mu\nu}\bar{\nabla}_\rho
 h_{\mu\nu}+\frac{2(\sigma +  \frac{\gamma  }{2 \mu^2\ell^2})}{
\ell^ 2}h^{\mu\nu}h_{\mu\nu}\\
&\hskip 4 cm -\frac{1}{\mu}\bar{\nabla}_\rho
h^{\mu\nu}\eta_\mu\,^{\rho\lambda}(\bar{\square}+\frac{2}
{\ell^ 2})h_{\lambda\nu}\bigg).\label{linaction}
\end{aligned}
\end{equation}
 The conjugate momenta for the field $h_{\mu \nu}$ are 
\begin{equation}
\Pi^{(1)\mu\nu}=-{\sqrt{-\bar{g}}\over64\pi G}
\left(\bar{\nabla}^0(2(\sigma +  \frac{\gamma  }{2 \mu^2\ell^2})h^{\mu\nu}+{1\over\mu}\eta^{\mu\rho}\,_{\lambda}
\bar{\nabla}_\rho
h^{\lambda \nu}) -{1\over\mu}\eta_\rho\,^{0\mu
}(\bar{\square}+{2\over \ell^ 2})h^{\rho\nu}\right),
\end{equation}
which, upon use of the field equations, yield 3 individual conjugate momenta for the corresponding degrees of freedom as   
\begin{equation}
\begin{aligned}
\Pi_M^{(1)\mu\nu}&=\frac{\sqrt{-\bar{g}}}{64\pi
G}\bigg(-(\sigma +  \frac{\gamma  }{2 \mu^2\ell^2})\bar{\nabla}^0h^{\mu\nu}+\frac{1}{\mu}(\mu^2-\frac{1}{
\ell^ 2})\eta_\rho\,^{0\mu}h_M^{\rho\nu}\bigg),\\
\Pi_L^{(1)\mu\nu}&=-{\sqrt{-\bar{g}}\over64\pi G}\bigg(2(\sigma +  \frac{\gamma  }{2 \mu^2\ell^2})-{1\over\mu
\ell}\bigg) \bar{\nabla}^0h_L^{\mu\nu},\\
\Pi_R^{(1)\mu\nu}&=-{\sqrt{-\bar{g}}\over64\pi G}\bigg(2(\sigma +  \frac{\gamma  }{2 \mu^2\ell^2})+{1\over\mu
\ell}\bigg)\bar{\nabla}^0h_R^{\mu\nu},
\end{aligned}
\end{equation}
where we define the right/left moving and the massive modes through the first order equations as
\begin{eqnarray}
	\left(\mathcal{D}^{L/R} h^{L/R} \right)_{\mu\nu}&=&h^{L/R}_{\mu\nu} \pm \ell \eta_{\mu}{}^{\alpha\beta} \nabla_\alpha h^{L/R}_{\beta\nu} \nonumber, \\
		\left(\mathcal{D}^{M} h^{M} \right)_{\mu\nu}&=&h^{M}_{\mu\nu} - \frac{1}{\mu\left(\sigma +  \frac{\gamma  }{2 \mu^2\ell^2}\right)} \eta_{\mu}{}^{\alpha\beta} \nabla_\alpha h^{M}_{\beta\nu}
\end{eqnarray}
Since we are dealing with a third order field theory, we need to employ the well-known Ostrogradsky procedure and introduce another canonical coordinate in the phase space,  which is the "time-derivative" of the spin-2 field defined as 
$K_{\mu\nu}\equiv\bar{\nabla}_0h_{\mu\nu}$ . The conjugate momenta of these new coordinates read as 
\begin{equation}
\Pi^{(2)\mu\nu}=\frac{-\sqrt{-\bar{g}}\bar{g}^{00}}{64\pi G\mu}\eta_\rho\,^{\lambda\mu}\bar{\nabla}_\lambda h^{\rho\nu},
\end{equation}
which once again split into 3 parts as 
\begin{equation}
\begin{aligned}
\Pi^{(2)\mu\nu}_M=\frac{-\sqrt{-\bar{g}}\bar{g}^{00}}{64\pi G}\bigg(\sigma +  \frac{\gamma  }{2 \mu^2\ell^2}\bigg)h_M^{\mu\nu},\hskip .5 cm
\Pi^{(2)\mu\nu}_R=\frac{\sqrt{-\bar{g}}\bar{g}^{00}}{64\pi G\mu
\ell}h_L^{\mu\nu},\hskip .5 cm
\Pi^{(2)\mu\nu}_L&=\frac{-\sqrt{-\bar{g}}\bar{g}^{00}}{64\pi
G\mu \ell}h_R^{\mu\nu}.
\end{aligned}
\end{equation}
Finally we have all the ingredients to write down the  Ostrogradsky Hamiltonian 
\begin{equation}
H \equiv \int d^2x\bigg(\dot{h}_{\mu\nu}\Pi^{(1)\mu\nu}+\dot{K}_{i\mu}\Pi^{(2)i\mu}- {\cal{L}}\bigg),
\end{equation}
from which follow the energies of the excitations for the left and right modes as ( note that the Lagrangian vanishes on-shell)
\begin{equation}
E_{L/R} = - \frac{c_{L/R}}{48\pi\ell}  \int d^2 x \, \sqrt{-\bar{g}} \,\bar{\nabla}^0  h_{L/R}^{\mu \nu} \, \partial_t h^{L/R}_{\mu \nu}, 
\label{energy11}
\end{equation}
while for the massive mode, the energy reads
\begin{equation}
E_{m_{g}^2} = \frac{m_{g}^2}{64\pi\mu G}   \int d^2 x \, \sqrt{-\bar{g}} \, \eta _\rho\,^{0\mu} h_{m_g}^{\rho \nu}\, \partial_t h^{m_g}_{\mu \nu}.\label{energy22}
\end{equation}
The solutions of the linearized equations can be classified using the representations of the algebra $SL(2,R) \times SL(2,R)$. As in the case of TMG, the solutions for the left/right moving modes correspond the weights $(2,0)$ and $(0,2)$ (see \cite{chiral} for the explicit form of the solutions). The weights for the massive mode are modified as
	\begin{equation}
h=\frac{3}{2}+\left(\sigma +  \frac{\gamma  }{2 \mu^2\ell^2}\right)\frac{ \mu \ell}{2}, \qquad \bar{h}= -\frac{1}{2}+\left(\sigma +  \frac{\gamma  }{2 \mu^2\ell^2}\right)\frac{ \mu \ell}{2}, 
\end{equation}
where the terms in parentheses should be positive for the positivity of the energy of the BTZ black hole, which is given by
\begin{equation}
	E_{BTZ} = \left(\sigma +  \frac{\gamma  }{2 \mu^2\ell^2}\right) M,
\end{equation}
up to a multiplicative constant. Using the explicit form of the solutions, one can show that the integrals in the energy expressions (\ref{energy11}-\ref{energy22}) are negative. Here let us depict the results for the primary fields only.  For the left and right modes one obtains a simple expression 
\begin{equation}
E_{L/R} =  \frac{c_{L/R}}{36  \ell}.  
\end{equation}
For the massive mode the result of the integral yields
\begin{equation}
E_{m_{g}^2} = \frac{m_{g}^2 \ell }{64\mu G} f(x) .\label{energy23}
\end{equation}
where  $x$ is a dimensionless parameter defined as
\begin{equation}
x \equiv  \left(\sigma +  \frac{\gamma  }{2 \mu^2\ell^2}\right)\frac{ \mu \ell}{2},
\end{equation}
and 
\begin{eqnarray}
f(x) &=&\frac{2^{4 x+5} (2 x+1)}{ 3+2x} \Bigg(- \frac{(3 +2 x)}{2 (x+1)} \, _2F_1 [2 (x+1),4 x+5;2 x+3;-1 ] \\ \nonumber
&& \qquad\qquad\qquad\qquad+ \,
   _2F_1 [2 x+3,4 x+5;2 (x+2);-1 ] \Bigg)
\end{eqnarray}
where the  $_2F_1$ functions are the hypergeometric functions.  The crucial point here is that  for the physical regions $x \in [0,  \infty) $ the function $f(x)$ is monotonically decreasing and takes the values as  $ f(x) \in [-1, -2)$ which yields  a negative energy for the massive mode for generic $m_g^2$. One obtains a similar result for the descendants. Therefore, our computation implies that  the bulk and boundary unitarity coexist only in the chiral limit, where $m_{g}^2=0$ and $c_L=0$. The drawback is that the perturbative log-modes arise at this point again.

\section{Exact Log-Modes}
As mentioned before, at the chiral point, there appear also {\it{exact}} log-modes. In order to study them, let us consider the following ansatz 
\begin{equation}
ds^2=\frac{\ell^2}{z^2}(dudv+dz^2)+F(u,z) du^2.
\end{equation}
Taking the trace of the field equations (\ref{mmg_denk}) gives a relation between the two parameters as
\begin{equation}
4\ell^2\Lambda_0+\frac{\gamma}{\ell^2\mu^2}+4\sigma=0,
\end{equation}
but the traceless part yields a third order differential equation \cite{Giribet}
\begin{equation}
p\,\partial_z^3 F(u,z)-\partial_z\bigg(\frac{\partial_z F(u,z)}{z}\bigg)=0,
\end{equation}
where we have defined
\begin{equation}
p\equiv \frac{2\ell\mu}{\gamma+2\ell^2\mu^2\sigma}=\frac{c_R-c_L}{c_R+c_L}.
\end{equation}
As long as $p\neq\pm1$, the generic solution is
\begin{equation}
F(u,z)= \frac{z^{1+\frac{1}{p}}}{1+\frac{1}{p}} c_1+c_2z^2+c_3.
\end{equation}
On the other hand, for $p=\pm 1$, the structure of the equation changes completely and one has the following solutions 
\begin{equation}
F(u,z)=f(u)\log z, \qquad \text{for} \,\, p=1 \,\, (c_L=0),
\end{equation}
and
\begin{equation}
F(u,z)=f(u)z^{2}\log z, \qquad \text{for} \,\, p=-1 \,\, (c_R=0),
\end{equation}
where $f(u)$ is a generic undetermined profile function of the null coordinate $u$. One can transform this solution to the global AdS coordinates using  the transformations given in \cite{Bayona}, but the resulting metric is highly complicated. Note also that these are the wave solutions of the MMG. Inserting these exact log-modes into the field equations of both TMG and NMG reveals that they are the solutions of these theories too, which has not been noticed in the literature before.

\section{Conclusions}
    To summarize the current situation regarding the bulk-boundary unitarity clash of $3D$ gravity theories, we note that  the only known theory free of this clash is the cosmological Einstein's  theory, which is locally trivial.  On the other hand, when one has local non-trivial  bulk dynamics, as in the case of  
	extended TMG, NMG, MMG, Born-Infeld gravity {\it{etc.}},  then a clash of bulk and boundary unitarity arises. We should note that our conclusion regarding MMG is only about its metric-formulation, where we integrated the linearized metric equations (\ref{lin_ark}) to arrive at the linearized action (\ref{linaction}), and it could be possible that MMG still survives in the first order formulation.~\footnote{We thank Eric Bergshoeff and Paul Townsend for this point.} Our results also have implications for possible MMG-like theories in higher dimensions. As noted in \cite{Bergshoeff:2015zga}, it might be possible to construct theories in higher dimensions whose field equations are consistent on-shell. However, it is not clear to us whether a non-geometric action as in the case of MMG is guaranteed to exist or not. Therefore, there might be an ambiguity in probing the physical properties of such theories because our results contradict with the result obtained by making use of the non-geometric action of the theory.

Since bartering triviality with  non-unitarity is not a good deal, one might worry if the current situation is an  
impasse and there simply does not exist a dynamical theory of gravity  which is unitary both in the bulk and on the boundary even in the simpler setting of  $3D$. The situation is actually not that bleak: there is still the possibility that the chiral gravity versions of  
the above-mentioned dynamical theories have linearization instability in $AdS$ and the problematic perturbative log-modes of \cite{Grumiller} are not integrable to full solutions: namely, these modes could be artifacts of the first order perturbation theory. Let us expound on this a little more.

In the case of pure GR, the linearization instability is known to occur for solutions with compact Cauchy surfaces \cite{Deser1973, Fischer1980} and spaces with Killing vectors. A similar phenomenon was also observed  in four dimensions in the pure quadratic gravity with the action
\begin{equation}
I =  -\frac{1}{4} \int d^4 x \sqrt{- g} \Big (  \alpha \,C_{\mu \nu \sigma \rho}   C^{\mu \nu \sigma \rho}    + \beta R^2 \Big ).
\end{equation}
When $ \alpha \beta \ge 0$, even though this theory admits linearized solutions with negative energy,  all exact solutions possess zero energy (see \cite{Boulware:1983} for the case of scale invariant gravity). Hence, there is a linearization instability about the flat space. In this theory, higher derivative terms require faster decaying conditions for the metric and this leads to an effective compactification of the Cauchy surface.   

There have been vigorous attempts \cite{Maloney,Carlip} to show the existence of the linearization instability in the chiral TMG. But, there are still some open questions regarding the choices of boundary conditions: namely, one imposes the Brown-Henneaux type boundary conditions to  obtain the linearization instability, but of course ideally one expects that out of all possible boundary conditions that yield the asymptotically $AdS$ space,  Brown-Henneaux conditions (not the weaker logarithmic ones) are forced  to eliminate the log-modes.  Moreover, if there is a linearization instability of $AdS$ in  chiral gravity, this does not only cast a question on the perturbative log modes but it also shows that all the perturbative treatment at the first order is 
in doubt. In this work, motivated by the earlier results obtained in the case of $4D$ pure quadratic gravity, we studied various higher derivative extensions of TMG: the $f(R)$-type modifications and MMG, which was previously proposed as a possible solution to the bulk-boundary unitarity conflict. We showed that, at the linearized level, these modifications only change the effective Newton's constant and one ends up with the same problems that occur in TMG. All these theories should be considered in the chiral limit and they allow the log-mode solutions with negative energy at the linearized level. However, at the nonlinear level, these theories are expected to behave differently, as far as the issue of the linearization instability is concerned, since they produce different higher-order terms which will constrain the linear perturbative result differently. Therefore, our results invite a systematic study of these theories. A detailed account of linearization instability will be given elsewhere \cite{AltasTekin}. 

\section{\label{ackno} Acknowledgements}
We would like to thank Eric Bergshoeff, Mehmet Ozkan, T.C. Sisman, Paul Townsend and Gideon Vos for useful discussions. B.T would like to thank the Groningen University for hospitality where major part of the work was carried and TUBITAK for support. E.K. is supported by the TUBITAK 2214-A Scholarship.  G.A. acknowledges support by a grant of the Dutch Academy of Sciences (KNAW). L.B is supported by the Dutch stichting voor Fundamenteel Onderzoek der Materie (FOM).

\end{document}